\documentstyle[12pt]{article}
\begin{document}
\begin{titlepage}
\begin{flushright}
Alberta Thy-1-95\\ January, 1995

\end{flushright}
\vskip1cm
\begin{center}
{\Large \bf Nonfactorization and the decays \\ $D_s^+ \rightarrow \phi \pi^+,
\phi \rho^+$
and $\phi l^+ \nu_l$  }\\[10mm]
A. N. Kamal and A. B. Santra\footnote{on leave of absence from Nuclear Physics
Division,
Bhabha Atomic Research Centre, Bombay 400085,  India.}\\[5mm]
{\em Theoretical Physics Institute and Department of Physics,\\ University of
Alberta,
Edmonton, Alberta T6G 2J1, Canada.}
\end{center}
\vskip1cm

\begin{abstract}
In six chosen scenarios for the $q^2$ dependence of the form factors involved
in $D_s^+
\rightarrow \phi $ transition, we have determined the allowed domain of  $x =
A_2(0) /
A_1(0)$ and $y = V(0)/A_1(0)$ from the experimentally measured ratios $R_{sl} =
\Gamma(D_s^+ \rightarrow \phi l^+ \nu_l)/\Gamma(D_s^+ \rightarrow \phi \pi^+)$
and  $R_h
= \Gamma(D_s^+ \rightarrow \phi \rho^+)/\Gamma(D_s^+ \rightarrow \phi \pi^+)$
in a
scheme that uses the $N_c =3$ value of the phenomenological parameter $a_1$ and
includes nonfactorized contribution. We find that the experimentally measured
values of $x$
and $y$ from semileptonic decays of $D_s^+$  favor solutions which have
significant
nonfactorized contribution, and, in particular, $R_{sl}$ favors solutions in
scenarios where
$A_1(q^2)$ is either flat or decreasing with $q^2$.
\end{abstract}
PACS index: 13.25Hw, 14.40.Nd
\end{titlepage}

\begin{center}
{\bf I. Introduction}
\end{center}

The phenomenological parameters $ a_1$ and $ a_2$, appearing in the description
of
charm-ed and bottom meson hadronic decays, are related, for $ N_c$ colors, to
the
perturbatively calculable Wilson coefficients $ C_1$ and $C_2$ by
\cite{rckl,wrbl},
\begin{eqnarray}
a_{1,2} = C_{1,2} + {C_{2,1} \over N_c}.
\end{eqnarray}

In the description of two-body hadronic decays of D mesons, phenomenology
seemed to
support the view that $ N_c \rightarrow  \infty$ limit applied \cite{bsw}. This
implied that $
a_1(m_c) \approx C_1(m_c) \approx 1.26$ and $ a_2(m_c) \approx C_2(m_c) \approx
-0.51$ \cite{nrsx}. However, the same idea ($ N_c \rightarrow  \infty$ limit)
carried over to
hadronic B decays failed. $ N_c \rightarrow  \infty$ limit would have implied $
a_1(m_b)
\approx 1.12$ and $ a_2(m_c)  \approx -0.26$ \cite{nrsx}, yet two-body B decay
data leave
no doubt  that $ a_2(m_b)$ is positive \cite{browder,cleo}.

We have recently proposed \cite{ks} that in $ B \rightarrow \psi (\psi(2S)) +
K(K^*)$ decays,
one ought to use $ N_c = 3$ values of  $ a_1$ and $ a_2$,  $ a_1(m_b) \approx
1.03$ and $
a_2(m_b) \approx 0.09$, and absorb the nonfactorized contributions in the
definition of
effective, and process-dependent, $ a_1$ and  $a_2$ for processes such as $ B
\rightarrow
PP, VP$ (P and V represent $ 0^-$ and $1^-$ mesons, respectively) which involve
only a
single Lorentz scalar. The idea of effective $ a_1$ and $ a_2$ is not new; it
has been
proposed and discussed by various authors in the past \cite{bs, hyc, cheng,
soares}. For
processes of kind $ B \rightarrow VV$, involving three independent Lorentz
structures (as
three partial waves, S, P,  and D, come into play), it is not possible to
define  effective $
a_1$  and $ a_2$ \cite{ks}. Pursuing this idea, we \cite{ks} had shown that
color-suppressed
decays $ B \rightarrow \psi (\psi(2S)) + K(K^*)$  and the longitudinal
polarization fraction in
$ B \rightarrow \psi  + K^*$ could be understood in all commonly used models
provided one
included a small nonfactorized amplitude, the amount depending on the model.
This
ameliorates the conundrum posed in \cite{gkp} where it was shown that none of
the
commonly used models explained the polarization data in $ B \rightarrow \psi  +
K^*$ decay
within the factorization assumption.

In the present paper we have carried this idea into the realm of charmed meson
decays; in
particular, we investigate the decays $D_s^+ \rightarrow \phi \pi^+, \phi
\rho^+$ and $\phi
l^+ \nu_l$.  Experimentally, only relative rates are measured (following the
notation
introduced in \cite{gkkp}),
\begin{eqnarray}
R_h \equiv {\Gamma(D_s^+ \rightarrow \phi \rho^+) \over \Gamma(D_s^+
\rightarrow \phi
\pi^+)}, \nonumber\\
R_{sl} \equiv {\Gamma(D_s^+ \rightarrow \phi l^+ \nu_l) \over \Gamma(D_s^+
\rightarrow
\phi \pi^+)}.
\end{eqnarray}

Though the problem has been discussed in \cite{gkkp}, we revisit it with the
purpose of
investigating the role of nonfactorized contribution in a formalism that uses
the $ N_c = 3$
values of $ a_1(m_c)$ and $ a_2(m_c)$. The reason for choosing the
Cabibbo-favored
decays of $ D_s^+$ is that the hadronic final states involve only a single
isospin and,
consequently, one does not have to worry about the interference effects due to
final state
interactions.

To the best of our knowledge nonfactorization contributions were first
discussed in charm
decays by Deshpande et al. \cite{ngd}.

In Section II, we describe the method and the calculation of the three decay
rates out of
which $ R_h$ and $ R_{sl}$ are constructed. The results are presented and
discussed in
Section III.

\begin{center}
{\bf II. Method and Calculations\\
A. Definitions}
\end{center}

The effective Hamiltonian for Cabibbo-favored hadronic charm decays is given by
\begin{eqnarray}
H_w = {G_F \over \sqrt{2} } V_{cs}V_{ud}^*\left\{ C_1 (\bar{u}d) (\bar{s}c) +
C_2 (\bar{u}c)
(\bar{s}d) \right\},
\end{eqnarray}
where $(\bar{u}d)$ etc. represent color-singlet (V - A) Dirac currents and
$C_1$ and $C_2$
are the Wilson coefficients for which we adopt the following values,
\begin{eqnarray}
C_1 = 1.26 \pm 0.04, \qquad C_2 = -0.51 \pm 0.05.
\end{eqnarray}
The central values of $C_1$ and $C_2$ are taken from \cite{nrsx}, though the
error
assignments are ours.

Fierz-transforming in color space with $N_c = 3$, we get
\begin{eqnarray}
(\bar{u}c)(\bar{s}d) = {1 \over 3} (\bar{u}d)(\bar{s}c) + {1 \over 2}
\sum_{a=1}^{8}{(\bar{u}\lambda^ac)(\bar{s}\lambda^ad)},
\end{eqnarray}
where $\lambda^a$ are the Gell-Mann matrices. We define, for $N_c = 3$,
\begin{eqnarray}
a_1 = C_1 + {C_2 \over 3} = 1.09 \pm 0.04, \nonumber\\
a_2 = C_2 + {C_1\over 3} = -0.09 \pm 0.05.
\end{eqnarray}
We also define \cite{bsw,nrsx} the relevant hadronic matrix elements and form
factors,
\begin{eqnarray}
\langle\rho^+|(\bar{u}d)|0\rangle&=&\varepsilon_\mu^* m_\rho f_\rho, \nonumber
\\
\langle\pi^+|(\bar{u}d)|0\rangle&=&-i f_{\pi} p_{\pi}^\mu, \nonumber \\
\langle\phi|(\bar{s}c)|D_s^+\rangle&=&i \left\{ (m_{D_s} + m_\phi)
\varepsilon_\mu^* A_1
(q^2) - {\varepsilon^*.q \over m_{D_s} + m_\phi} (p_{D_s} + p_\phi)_\mu A_2
(q^2) \right.
\nonumber \\
&-& \left.2 m_\phi {\varepsilon^*.q \over q^2}  q_\mu A_3(q^2)+
{\varepsilon^*.q \over q^2}
(2m_\phi)q_\mu A_0(q^2) \right\}  \nonumber \\
&+& {2 \over m_{D_s} + m_\phi}\varepsilon_{\mu \nu \rho
\sigma}\varepsilon^{*\nu}p^\rho_{D_s} p^\sigma_\phi V(q^2),
\end{eqnarray}
where $q_\mu = (p_{D_s} - p_\phi)_\mu$. In addition, the following constraints
apply at $q^2
= 0$,
\begin{eqnarray}
A_0(0)&=&A_3(0), \nonumber \\
2 m_\phi A_3(0)&=&(m_{D_s} + m_\phi) A_1(0) - (m_{D_s} - m_\phi) A_2(0).
\end{eqnarray}
With these preliminaries, we now study the rates for the processes $D_s^+
\rightarrow  \phi
l^+ \nu_l$,  $\phi \pi^+$ and $\phi \rho^+$.

\begin{center}
{\bf  B. $D_s^+ \rightarrow  \phi l^+ \nu_l$ decay}
\end{center}

In the notation of  \cite{gkkp},
\begin{eqnarray}
{d\Gamma \over dt^2} \left( D_s^+ \rightarrow  \phi l^+ \nu_l \right) = {G_F^2
m_{D_s}^5
\over 192 \pi^3} |V_{cs}|^2 k(t^2)\sum_{\lambda}^{}{H_{\lambda \lambda}(t^2)},
\end{eqnarray}
where, $H_{\lambda\lambda}(t^2)$ in each helicity state are defined by,
\begin{eqnarray}
H_{00}(t^2)&=&\left( {1 + r \over 2r} \right)^2 \biggl | (1 - r^2 -t^2)
A_1(t^2) - {k^2(t^2) \over (1
+r)^2}A_2(t^2)\biggr |^2, \nonumber \\
H_{\pm\pm}(t^2)&=&(1 + r)^2 t^2\biggl | A_1(t^2) \mp {k(t^2) \over (1
+r)^2}V(t^2)\biggr |^2,
\end{eqnarray}
with
\begin{eqnarray}
r = {m_\phi \over m_{D_s}}, \qquad t^2 = {q^2 \over m_{D_s}^2}, \qquad 0 \leq
t^2 \leq (1
-r)^2.
\end{eqnarray}
In the rest frame of $D_s^+$, the magnitude of $\phi$ meson 3-momentum is given
by
\begin{eqnarray}
p(t^2)&=&{m_{D_s} \over 2} k(t^2), \nonumber \\
\mbox{with} \qquad
k(t^2)&=&[(1 + r^2 - t^2)^2 - 4r^2]^{1/2}.
\end{eqnarray}

\begin{center}
{\bf  C. $D_s^+ \rightarrow  \phi \pi^+$ decay}
\end{center}

Using (3), (5) and (6), the decay amplitude for $D_s^+ \rightarrow  \phi \pi^+$
is written as
\begin{eqnarray}
A(D_s^+ \rightarrow  \phi \pi^+) = {G_F \over \sqrt{2} } V_{cs} V_{ud}^*\left\{
a_1 \left\langle
\phi \pi^+|(\bar{u}d)(\bar{s}c)|D_s^+ \right\rangle + C_2  \left\langle \phi
\pi^+|H_w^{(8)}|D_s^+ \right\rangle\right\},
\end{eqnarray}
where, from (5)
\begin{eqnarray}
H_w^{(8)} \equiv {1 \over 2}
\sum_{a}^{}{(\bar{u}\lambda^ac)(\bar{s}\lambda^ad)}.
\end{eqnarray}

We calculate the first term in the brackets of (13) in the factorization
approximation using the
definitions in (6) and parametrize the second, the nonfactorized term for
convenience of
combining it with the first term, as follows,
\begin{eqnarray}
\left\langle \phi \pi^+|H_w^{(8)}|D_s^+ \right\rangle = 2m_\phi f_{\pi}
\varepsilon^*.p_{D_s}
A_0^{NF}(m_{\pi}^2).
\end{eqnarray}
This nonfactorized parameter, $A_0^{NF}$, represents a spurion scattering
$H_w^{(8)} +
D_s^+ \rightarrow \phi + \pi^+$, at the Mandelstam point $s = m_{D_s}^2$, $t =
m_{\pi}^2$
and $u = m_{\phi}^2$.

Combining the factorized and the nonfactorized terms we obtain,
\begin{eqnarray}
A(D_s^+ \rightarrow \phi \pi^+)&=&\sqrt{2} G_F V_{cs}V_{ud}^*m_\phi a_1^{eff}
A_0(m_{\pi}^2) \varepsilon^*.p_{D_s},
\end{eqnarray}
where,
\begin{eqnarray}
a_1^{eff}&=&a_1\biggl\{ 1 + {C_2 \over a_1}  { A_0^{NF}(m_{\pi}^2)\over
A_0(m_{\pi}^2)}
\biggr\}.
\end{eqnarray}

In terms of the quantities defined in (10) - (12), the decay rate $D_s^+
\rightarrow \phi \pi^+
$ is given by \cite{gkkp},
\begin{eqnarray}
\Gamma\left( D_s^+ \rightarrow \phi \pi^+ \right) = {G_F^2 m_{D_s}^5\over 32
\pi} |V_{cs}|^2
|V_{ud}|^2 |a_1^{eff}|^2 \left( {f_{\pi} \over m_{D_s}} \right)^2 k(0)
H_{00}(0),
\end{eqnarray}
where we have approximated $m_\pi^2/m_{D_s}^2 \approx 0$ and
\begin{eqnarray}
k(0)H_{00}(0) = (1 - r^2)^3 \left( {1 + r \over 2r} \right)^2 \biggl | A_1(0) -
{1 - r \over 1 + r}
A_2(0)\biggr |^2.
\end{eqnarray}

We choose not to work with the form (18) which involves $A_0(0)$ in the
definition of
$a_1^{eff}$ but rather with a form in which $A_0(0)$ is eliminated altogether
in favor of
$A_1(0)$ and $A_2(0)$ via  (8). In doing so, we obtain,
\begin{eqnarray}
\Gamma\left( D_s^+ \rightarrow \phi \pi^+ \right) = {G_F^2 m_{D_s}^5\over 32
\pi} |V_{cs}|^2
|V_{ud}|^2 a_1^2 \left( {f_{\pi} \over m_{D_s}} \right)^2 k(0) \hat{H}_{00}(0),
\end{eqnarray}
where,
\begin{eqnarray}
k(0)\hat{H}_{00}(0) = (1 - r^2)^3 \left( {1 + r \over 2r} \right)^2 \biggl |
A_1(0) - {1 - r \over 1 +
r} A_2(0) + {2r \over 1 + r} {C_2 \over a_1} A_0^{NF}(0) \biggr |^2.
\end{eqnarray}

\begin{center}
{\bf  D. $D_s^+ \rightarrow  \phi \rho^+$ decay with a zero width $\rho$ meson}
\end{center}

{}From (3), (5) and (6), the decay amplitude for $D_s^+ \rightarrow  \phi
\rho^+$ is written as,
\begin{eqnarray}
A\left( D_s^+ \rightarrow  \phi \rho^+ \right) = {G_F \over \sqrt{2} } V_{cs}
V_{ud}^* \left\{ a_1
\left\langle \phi \rho^+ |(\bar{u}d)(\bar{s}c)|D_s^+ \right\rangle + C_2
\left\langle \phi \rho^+|
H_w^{(8)} |D_s^+\right\rangle\right\},
\end{eqnarray}
where $H_w^{(8)}$ is defined in (14).

Again we calculate the first term in the brackets of (22) in the factorization
approximation
and define the nonfactorized second term, for ease of combining it with the
factorized first
term, as follows,
\begin{eqnarray}
\left\langle \phi \rho^+| H_w^{(8)} |D_s^+\right\rangle&=&i m_{\rho} f_{\rho}
\biggl\{
(m_{D_s} + m_\phi ) (\varepsilon^*_{\rho}. \varepsilon^*_{\phi})
A_1^{NF}(m_{\rho}^2)
\nonumber \\
&-&{2 \over m_{D_s} + m_\phi} (\varepsilon^*_{\rho}.p_{D_s})
(\varepsilon^*_{\phi}.p_{D_s})
A_2^{NF}(m_{\rho}^2)  \nonumber \\
&-&  {2i \over m_{D_s} + m_\phi} \varepsilon_{\mu\nu\rho\sigma}
\varepsilon_{\rho}^{*\mu}
\varepsilon_{\phi}^{*\nu} p_{D_s} ^\rho p_\phi^\sigma V^{NF}(m_{\rho}^2)
\biggr\}.
\end{eqnarray}
Further, because of lack of phase space in this decay mode, we retain
nonfactorized
contribution only to S-waves, that is,
\begin{eqnarray}
A_1^{NF} (m_{\rho}^2)&\neq&0, \nonumber \\
A_2^{NF} (m_{\rho}^2)&=&V^{NF} (m_{\rho}^2) = 0.
\end{eqnarray}
The decay amplitudes in each helicity state are then given by
\begin{eqnarray}
A_{00}(D_s^+ \rightarrow  \phi \rho^+)&=&{G_F \over \sqrt{2} } V_{cs} V_{ud}^*
m_{\rho}
f_{\rho} (m_{D_s} + m_\phi) a_1 \left\{ a \xi A_1 (m_{\rho}^2) - b A_2
(m_{\rho}^2)
\right\},\nonumber \\
A_{\pm\pm}(D_s^+ \rightarrow  \phi \rho^+)&=&{G_F \over \sqrt{2} } V_{cs}
V_{ud}^*
m_{\rho} f_{\rho} (m_{D_s} + m_\phi) a_1 \left\{  \xi A_1(m_{\rho}^2) \mp c
V(m_{\rho}^2)
\right\},
\end{eqnarray}
where, with $t_\rho = m_{\rho}/m_{D_s}$,
\begin{eqnarray}
a&=&{m_{D_s}^2 - m_\phi^2 - m_{\rho}^2 \over 2m_{\rho}m_\phi} = {1 - r^2 -
t_\rho^2 \over
2rt_\rho}, \nonumber \\
b&=&{2|\vec{p}_\phi|^2 m_{D_s}^2  \over (m_{D_s} + m_\phi)^2 m_{\rho}m_\phi} =
{k^2(t_\rho^2) \over 2rt_\rho(1 + r)^2}, \nonumber \\
c&=&{2|\vec{p}_\phi| m_{D_s} \over (m_{D_s} + m_\phi)^2 } \; = \; {k(t_\rho^2)
\over (1 +
r)^2},  \\
\xi&=&1 + {C_2 \over a_1} \chi_\rho, \nonumber \\
\chi_\rho&=&{A_1^{NF} (m_{\rho}^2)\over A_1 (m_{\rho}^2)}. \nonumber
\end{eqnarray}

The decay rate can then be expressed in a form resembling the expression given
in
\cite{gkkp},
\begin{eqnarray}
\Gamma\left( D_s^+ \rightarrow \phi \rho^+ \right) = {G_F^2 m_{D_s}^5\over 32
\pi}
|V_{cs}|^2 |V_{ud}|^2 a_1^2 \left( {f_{\rho} \over m_{D_s}} \right)^2
k(t_\rho^2)
\sum_{\lambda}^{}{H_{\lambda\lambda}^\prime(t_\rho^2)},
\end{eqnarray}
with
\begin{eqnarray}
H_{00}^\prime(t_\rho^2)&=&\left( {1 + r \over 2r} \right)^2 \biggl | (1 - r^2 -
t_\rho^2) \xi A_1
(t_\rho^2) - {k^2(t_\rho^2) \over (1 + r)^2} A_2 (t_\rho^2)   \biggr |^2,
\nonumber \\
H_{\pm\pm}^\prime(t_\rho^2)&=&(1 + r )^2 t_\rho^2 \biggl |  \xi A_1 (t_\rho^2)
 \mp
{k(t_\rho^2) \over (1 + r)^2} V(t_\rho^2)   \biggr |^2.
\end{eqnarray}

\begin{center}
{\bf  E. $D_s^+ \rightarrow  \phi \rho^+$ decay with a finite width $\rho$
meson.}
\end{center}

The finite width of $\rho$ meson is taken into account by smearing the rates
given in (27)
and (28) over the $\rho$ width by using a unit normalized Breit-Wigner measure,
$\rho(t^2)$.
This is accomplished by the replacement \cite{gkkp,pv,kv},
\begin{eqnarray}
k(t_\rho^2) H_{\lambda\lambda}(t_\rho^2) \rightarrow \int_{4t_\pi^2}^{(1 -
r)^2}{k(t^2)H_{\lambda\lambda}(t^2) \rho(t^2)dt^2}
\end{eqnarray}
with
\begin{eqnarray}
\int_{4t_\pi^2}^{\infty}{\rho(t^2)dt^2} = 1,  \qquad  \biggl ( t_\pi = {m_\pi
\over m_{D_s}} \biggr
)
\end{eqnarray}
Two measures that have been used in the past are \cite{pv,kv},
\begin{eqnarray}
\rho_1(t^2) = {N_1 \over \pi} {\gamma_\rho t_\rho \over (t^2 - t_\rho^2)^2 +
\gamma_\rho^2
t_\rho^2},
\end{eqnarray}
and
\begin{eqnarray}
\rho_2(t^2) = {N_2 \over \pi} {\gamma(t^2) t_\rho \over (t^2 - t_\rho^2)^2 +
\gamma^2(t^2)
t_\rho^2},
\end{eqnarray}
with
\begin{eqnarray}
{1 \over N_{1,2}} = \int_{4t_\pi^2}^{\infty}{\rho_{1,2}(t^2)dt^2}.
\end{eqnarray}
The appearance of $N_{1,2}$ in (31) and (32) ensures that $\rho_{1,2}(t^2)$ are
unit
normalized. In  (31), $\gamma_\rho = \Gamma_\rho/m_{D_s^+}$, where
$\Gamma_\rho$ is
the $\rho$ width. In (32), $\gamma(t^2)$ is so chosen as to reflect the P-wave
nature of the
$\rho$ meson
\begin{eqnarray}
\gamma(t^2) = \gamma_\rho {t_\rho^2 \over t^2} \left[ {t^2 - 4t_\pi^2 \over
t_\rho^2 - 4
t_\pi^2}\right]^{3/2}.
\end{eqnarray}
For a $\rho$ width of 151.2 MeV and $m_\rho $ = 769.1 MeV, we found $N_1$ =
1.0758
and  $N_2$ = 0.9946. In our calculation we have used the smearing function
$\rho_2(t^2)$
with energy-dependent width.

\begin{center}
{\bf  III.  Results and Discussions}
\end{center}

In the results presented below we have used $V_{cs} = V_{ud} = 0.975$ and only
the
central values of $a_1$, $a_2$ and $C_2$ given in (4) and (6): $ a_1$ = 1.09,
$a_2$ =
-0.09 and $C_2$ = -0.51. Defining
\begin{eqnarray}
x  = {A_2(0) \over A_1(0)}, \qquad y  = {V(0) \over A_1(0)},
\end{eqnarray}
we can write the decay rate for $D_s^+ \rightarrow \phi \pi^+$ as
\begin{eqnarray}
\Gamma(D_s^+ \rightarrow \phi \pi^+) = 0.2341 \times10^{12} |A_1(0)|^2 \left(
1 - 0.3177 x
- 0.3192 \chi_\pi \right) ^2 s^{-1},
\end{eqnarray}
where,
\begin{eqnarray}
 \chi_\pi = {A_0^{NF} \over A_1(0)} .
\end{eqnarray}

We calculate $\Gamma(D_s^+ \rightarrow \phi l^+\nu_l)$ and the smeared
$\Gamma(D_s^+
\rightarrow \phi \rho^+)$ in six different scenarios listed below.
\vskip  .2cm
\noindent {\bf Scenario 1:}  All form factors extrapolate in $q^2$ as
monopoles; $A_1(q^2)$
and $A_2(q^2)$ with pole mass 2.53 GeV and $V(q^2)$ with a pole mass 2.11 GeV
(BSWI
\cite{bsw} ). We get
\begin{eqnarray}
\Gamma(D_s^+ \rightarrow \phi l^+ \nu_l) = 0.1615 \times 10^{12}
|A_1(0)|^2\left ( 1 - 0.2840
x + 0.0344 x^2 + 0.0144 y^2 \right ) s^{-1}, \nonumber \\
\Gamma(D_s^+ \rightarrow \phi \rho^+) = 0.3972 \times 10^{12} |A_1(0)|^2\left (
\xi^2 -
0.1493 x \xi+0.0118 x^2 + 0.0167 y^2 \right ) s^{-1}.
\end{eqnarray}
\vskip  .2cm
\noindent {\bf Scenario 2:} $A_1(q^2)$ extrapolates in $q^2$ as monopole with
pole mass
2.53 GeV and $A_2(q^2)$ and $V(q^2)$ as dipoles with pole masses 2.53 and 2.11
GeV,
respectively (BSWII \cite{nrsx} ). We obtain
\begin{eqnarray}
\Gamma(D_s^+ \rightarrow \phi l^+ \nu_l) = 0.1615 \times 10^{12}
|A_1(0)|^2\left ( 1 - 0.2958
x + 0.0367 x^2 + 0.0176 y^2 \right ) s^{-1}, \nonumber \\
\Gamma(D_s^+ \rightarrow \phi \rho^+) = 0.3972 \times 10^{12} |A_1(0)|^2\left (
\xi^2 -
0.1623 x \xi +0.0137 x^2 + 0.0218 y^2 \right ) s^{-1}.
\end{eqnarray}
\vskip  .2cm
\noindent {\bf Scenario 3:} $A_1(q^2)$ flat, $A_2(q^2)$ and $V(q^2)$
extrapolate in $q^2$
as monopoles with pole masses 2.53 and 2.11 GeV, respectively. We obtain
\begin{eqnarray}
\Gamma(D_s^+ \rightarrow \phi l^+ \nu_l) = 0.1423 \times 10^{12}
|A_1(0)|^2\left ( 1 - 0.3098
x + 0.0390 x^2 + 0.0163 y^2 \right ) s^{-1}, \nonumber \\
\Gamma(D_s^+ \rightarrow \phi \rho^+) = 0.3297 \times 10^{12} |A_1(0)|^2\left (
\xi^2 -
0.1656 x \xi +0.0142 x^2 + 0.0202 y^2 \right ) s^{-1}.
\end{eqnarray}
\vskip  .2cm
\noindent {\bf Scenario 4:} $A_1(q^2)$ decreasing linearly in $q^2$,
\begin{eqnarray}
A_1(q^2) = A_1(0)\left\{ 1 - \left( {q \over 2.53} \right)^2 \right\},
\end{eqnarray}
and  $A_2(q^2)$ and $V(q^2)$ extrapolate in $q^2$ as monopoles with pole masses
2.53
and 2.11 GeV, respectively. This yields
\begin{eqnarray}
\Gamma(D_s^+ \rightarrow \phi l^+ \nu_l) = 0.1262 \times 10^{12}
|A_1(0)|^2\left ( 1 - 0.3361
x + 0.0440 x^2 + 0.0184 y^2 \right ) s^{-1}, \nonumber \\
\Gamma(D_s^+ \rightarrow \phi \rho^+) = 0.2742 \times 10^{12} |A_1(0)|^2\left (
\xi^2 -
0.1835 x \xi +0.0171 x^2 + 0.0243 y^2 \right ) s^{-1}.
\end{eqnarray}
\vskip  .2cm
\noindent {\bf Scenario 5:} $A_1(q^2)$ and $A_2(q^2)$ flat and $V(q^2)$
extrapolates in
$q^2$ as monopole with pole mass 2.11 GeV \cite{gkkp}. We get
\begin{eqnarray}
\Gamma(D_s^+ \rightarrow \phi l^+ \nu_l) = 0.1423 \times 10^{12}
|A_1(0)|^2\left ( 1 - 0.2980
x + 0.0366 x^2 + 0.0163 y^2 \right ) s^{-1}, \nonumber \\
\Gamma(D_s^+ \rightarrow \phi \rho^+) = 0.3297 \times 10^{12} |A_1(0)|^2\left (
\xi^2 -
0.1526 x \xi +0.0122 x^2 + 0.0202 y^2 \right ) s^{-1}.
\end{eqnarray}
\vskip  .2cm
\noindent {\bf Scenario 6:} $A_1(q^2)$ decreasing linearly in $q^2$ according
to
\cite{gkkp},
\begin{eqnarray}
A_1(q^2) = A_1(0)\left\{ 1 - \left( {q \over 3.5} \right)^2 \right\},
\end{eqnarray}
and  $A_2(q^2)$ and $V(q^2)$ extrapolate in $q^2$ as monopoles with pole masses
2.53
and 2.11 GeV, respectively. This yields
\begin{eqnarray}
\Gamma(D_s^+ \rightarrow \phi l^+ \nu_l) = 0.1337 \times 10^{12}
|A_1(0)|^2\left ( 1 - 0.3172
x + 0.0403 x^2 + 0.0174 y^2 \right ) s^{-1}, \nonumber \\
\Gamma(D_s^+ \rightarrow \phi \rho^+) = 0.3000 \times 10^{12} |A_1(0)|^2\left (
\xi^2 -
0.1667 x \xi +0.0145 x^2 + 0.0222 y^2 \right ) s^{-1}.
\end{eqnarray}

For data, we use
\begin{eqnarray}
R_h&=&1.86 \pm 0.26^{+0.29}_{-0.40}\; , \qquad \cite{pav1} \\
R_{sl}&=&0.54 \pm 0.10 \; . \qquad \qquad\cite{msw}
\end{eqnarray}
In the following, we discuss the analysis of the ratios $R_{sl}$ and $R_{h}$
separately.
\vskip  .2cm
\noindent {\bf $R_{sl}$ :} We reiterate that we are using the $N_c = 3$ value
of $a_1 = 1.09$
(central value only). $R_{sl }$ is constructed from (36) for $\Gamma(D_s^+
\rightarrow \phi
\pi^+)$ and (38)-(45) for $\Gamma(D_s^+ \rightarrow \phi l^+ \nu_l)$ in various
scenarios.
 The allowed region in $(x,y)$ plane are shown in Fig.1 for some selected
scenarios as
explained below. Firstly, we observe that no solutions were found for $\chi_\pi
= 0$ (see
(37)) in scenarios 1 and 2 (BSWI and BSWII, respectively). In contrast,
solutions were found
in  these scenarios in \cite{gkkp}; the difference lies in our use of $a_1 =
1.09$ while  $a_1
\approx 1.26$ was used in \cite{gkkp}. Solutions, however, were found for
$\chi_\pi = 0$ in
all the other four scenarios. Secondly, the scenarios for $A_1(q^2)$ flat, and
$A_1(q^2)$
and $A_2(q^2)$ flat, for $\chi_\pi = 0$ or $\chi_\pi  \neq 0$ were almost
indistinguishable,
while those for BSWI and BSWII with $\chi_\pi \neq 0$ were very similar.
Consequently, we
have chosen to plot only the results using  scenarios 1, 3 and 4 to keep Fig.1
uncluttered.
The experimental points are from Refs. \cite{kk}, \cite{plf} and \cite{pav2}
and the plots are
made for $\chi_\pi = 0$ and -0.5.

The absolute rate, $\Gamma(D_s^+ \rightarrow \phi \pi^+)$ and the branching
ratio,
$B(D_s^+ \rightarrow \phi \pi^+)$ can not  be calculated in a model-independent
way, but,
for the record, in BSW model \cite{bsw}, one gets
\begin{eqnarray}
B(D_s^+ \rightarrow \phi \pi^+)&=&2.48\; \% \qquad (\chi_\pi = 0),\nonumber \\
&=&4.03\; \% \qquad (\chi_\pi = -0.5).
\end{eqnarray}
Particle Data Group \cite{lm} list $B(D_s^+ \rightarrow \phi \pi^+) = (3.5 \pm
0.4) \%$,
though a direct measurement of the branching ratio does not exist.

Clearly from Fig.1 one notes that data prefer solutions with $A_1(q^2)$ flat or
$A_1(q^2)$
decreasing with $q^2$ (scenarios 3 to 6). Further, solutions with a
nonfactorized
contribution, $\chi_\pi = -0.5$, fare much better than those with $\chi_\pi =
0$. In particular,
E-687 data \cite{plf} are consistent with all the six scenarios with $\chi_\pi
= -0.5$. CLEO
data  \cite{pav2} are consistent with scenarios 3 to 6 ($A_1(q^2)$ flat or
decreasing with
$q^2$) with $\chi_\pi = -0.5$. E-653 data \cite{kk} do not admit a solution
with $-0.5 \leq
\chi_\pi \leq 0$.

\vskip  .2cm
\noindent {\bf $R_{h}$ :} The ratio $R_h$, eq.(2), is constructed from the
rates
$\Gamma(D_s^+ \rightarrow \phi \pi^+)$ given in (36) and $\Gamma(D_s^+
\rightarrow \phi
\rho^+)$, for finite width $\rho$ meson, given in (38) to (45). The allowed
regions in $(x,y)$
plane for  the six scenarios are shown in Figs. 2, 3 and 4. We find that with
$\chi_\pi =
\chi_\rho = 0$, solutions accommodate E-687 data \cite{plf} in all scenarios,
while CLEO
data \cite{pav2} are accommodated in scenarios 4 and 6 ($A_1(q^2)$ decreasing
with
$q^2$).

However with $\chi_\pi  = -0.5$ and $\chi_\rho = 0.5$, all three data points
\cite{kk,plf,pav2}
are accommodated in all six scenarios. The allowed region in the $(x,y)$ plane
is now a
band with an upper and lower branch.

For the record, with $\chi_\rho = 0.5$, and using BSWII model  \cite{nrsx}, one
gets
\begin{eqnarray}
B(D_s^+ \rightarrow \phi \rho^+) = 6.23 \; \%.
\end{eqnarray}
Again, though this branching ratio is not directly measured, Particle Data
Group \cite{lm} list
\begin{eqnarray}
B(D_s^+ \rightarrow \phi \rho^+) = (6.5^{+1.6}_{-1.8}) \; \%
\end{eqnarray}

In conclusion, taking the $N_c = 3$ value of $a_1$ seriously, we have asked:
What is the
domain of $x$ and $y$ allowed by the ratios $R_h$ and $R_{sl}$ in six chosen
scenarios for
the $q^2$ dependence of the form factors, with and without nonfactorization
contribution?
And,  what is the size and the sign of the nonfactorization contribution in
order that the
measured values of  $x$ and  $y$ fall within the allowed domain obtained from
$R_h$ and
$R_{sl}$? We should emphasize that the experimental determination of $x$ and
$y$ from
semileptonic data is not model free as monopole extrapolation is assumed for
all the form
factors in data analysis. We find that an analysis with the inclusion of
nonfactorized
contribution fares much better in selecting an allowed domain of $x$ and $y$
consistent
with the data. This is particularly true of $R_{sl}$ where, in addition, the
scenarios in which
$A_1(q^2)$ is flat, or decreasing with $q^2$ are favored over the scenarios
where it rises
with $q^2$. As to the size and sign of the nonfactorized contribution, we have
no
explanation.

\begin{center}
{\bf  Acknowledgements}
\end{center}

ANK wishes to acknowledge a research grant from the Natural Sciences and
Engineering
Research Council of Canada which partially supported this research.

\newpage
\noindent {\bf Figure Captions}
\vskip 3mm
\noindent  {\bf Fig.1:} The regions enclosed by various curves and the axes are
the domains
of ($ x,y$) allowed by $ R_{sl}$ (eqn. (47)) in different scenarios. The solid
curve is for
scenario 1 with $\chi_\pi = -0.5$; the dashed (second from the right) and the
dash-dashed
(innermost) are for scenario 3 with $\chi_\pi = -0.5$ and 0, respectively; the
dotted
(outermost) and the dot -dashed (second from left) are for scenario 4 with
$\chi_\pi = -0.5$
and 0, respectively.  The corresponding curves for scenario 6 (not shown to
avoid
cluttering) lie between the curves for scenarios 3 and 4. Also shown are the
data points; A:
E-687  \cite{plf}, B: E-653 \cite{kk} and C: CLEO \cite{pav1}.
\vskip 3mm
\noindent  {\bf Fig.2:} The domain (between two solid lines) of  ($ x,y$)
allowed by $R_h$
(eqn. (46)) with $ \chi_\rho = 0.5$ and $ \chi_\pi = -0.5$. (a) in scenario 1;
(b) in scenario 2.
The region enclosed by the dashed lines and the axes both in (a) and (b) is the
allowed
domain of ($x,y$) with $\chi_\rho = \chi_\pi = 0$. Also shown are the data
points; A: E-687
\cite{plf}, B: E-653 \cite{kk} and C: CLEO \cite{pav1}.
\vskip 3mm
\noindent  {\bf Fig.3:} Same as Fig. 2. (a) in scenario 3; (b) in scenario 5.

\vskip 3mm
\noindent  {\bf Fig.4:} Same as Fig. 2. (a) in scenario 4; (b) in scenario 6.


\newpage

\begin{thebibliography}{99}

\bibitem{rckl}
R. R\H{u}ckl, Habilitationsschrift, University of Munich, 1983.

\bibitem{wrbl}
M. Wirbel, Prog. Part. Nucl. Phys. {\bf 21}, 33 (1988)

\bibitem{bsw}
M. Bauer, B. Stech and M. Wirbel, Z. Phys. C {\bf 34}, 103 (1987); see also,
M. Wirbel, B. Stech and M. Bauer, Z. Phys. C {\bf29},  637 (1985)
M. Bauer and M. Wirbel, Z. Phys. C {\bf42},  671 (1989)

\bibitem{nrsx}
M. Neubert, V. Rieckert, B.Stech and Q. P. Xu in Heavy Flavours,  ed. by  A. J.
Buras and
M. Lindner, World Scientific, Singapore, 1992.

\bibitem{browder}
T. E. Browder, K. Honscheid and S. Playfer, Report CLNS 93/1261, to appear in B
Decays,
$2^{ \em nd}$ Edition, ed. by S. Stone. World Scientific, Singapore.

\bibitem{cleo}
M. S. Alam et al. ( CLEO collaboration) Phys. Rev. {\bf D50},43 (1994).

\bibitem{ks}
A. N. Kamal and A. B. Santra, University of Alberta Report No. Alberta-Thy
31-94.

\bibitem{bs}
B. Blok and M. Shifman, Nucl. Phys. {\bf B399}, 441, 459 (1993), ibid {\bf
B389}, 534 (1993)
and references therein.

\bibitem{hyc}
H. -Y Cheng, Taipei Report No. IP-ASTP-11-94.

\bibitem{cheng}
H. -Y. Cheng and B. Tseng, Taipei Report No. IP-ASTP-21-94.

\bibitem{soares}
J. M. Soares, TRUIMF Report No. TRI-PP-94-78.

\bibitem{gkp}
M. Gourdin, A. N. Kamal and X. Y. Pham, Paris Report No. PAR/LPTHE/94-19, to
appear in
Phys. Rev. Lett.

\bibitem{gkkp}
M. Gourdin, A. N. Kamal, Y. Y. Keum and X. Y. Pham, Phys. Lett {\bf B339} 173
(1994).

\bibitem{ngd}
N. Deshpande, M. Gronau and D. Sutherland, Phys.  Lett. {\bf 90B},  431  (1980)
{}.

\bibitem{pv}
X. Y. Pham and X. C. Vu, Phys. Rev. {\bf D46}, 261 (1992)

\bibitem{kv}
A. N. Kamal and R. C. Verma, Phys. Rev. {\bf D45}, 982 (1992)

\bibitem{pav1}
P. Avery et al. CLEO collaboration, Phys. Rev. Lett. {\bf 68}, 1279 (1992)

\bibitem{msw}
M. S. Witherell, in Lepton Photon Interaction, XVI International Symposium,
Ithaca, N. Y.
1993, ed. by P. Drell and D. Rubin, AIP conference Proceedings 302, p198,
American
Institute of Physics, New York (1994).


\bibitem{kk}
K. Kodama et al., E-653 Collaboration, Phys. Lett.  {\bf B309}, 483 (1993).

\bibitem{plf}
P. L. Frabetti et al., E-687 Collaboration, Phys. Lett. {\bf B328}, 187 (1994).

\bibitem{pav2}
P. Avery et al., CLEO Collaboration, Phys. Lett. {\bf B337}, 405 (1994).

\bibitem{lm}
L. Montanet et al., Particle Data Group, Review of Particle Properties, Phys.
Rev. {\bf D50},
1173 (1994).

\end{thebibliography}
\end{document}